\documentclass[aps,pra,preprintnumbers,showpacs,tightenlines]{revtex4}

\usepackage{amssymb}
\usepackage{amsmath}
\usepackage{graphicx}
\usepackage{epsfig}
\usepackage{subfigure}
\usepackage{amsfonts}
\usepackage{CJK}

\begin{document}

\title{Generating entanglement between microwave photons and qubits in multiple cavities
coupled by a superconducting qutrit}

\author{Chui-Ping Yang$^{1,4}$, Qi-Ping Su$^{1}$, Shi-Biao Zheng$^{2}$, and Siyuan Han$^{3}$}

\address{$^1$Department of Physics, Hangzhou Normal University,
Hangzhou, Zhejiang 310036, China}

\address{$^2$Department of Physics, Fuzhou University, Fuzhou 350002,
China}

\address{$^3$Department of Physics and Astronomy, University of
Kansas, Lawrence, Kansas 66045, USA}

\address{$^4$State Key Laboratory of Precision Spectroscopy, Department of Physics,
East China Normal University, Shanghai 200062, China}

\date{\today}

\begin{abstract}
We discuss how to generate entangled coherent states of
four \textrm{microwave} resonators \textrm{(a.k.a. cavities)}
coupled by a superconducting qubit. We also show \textrm{that} a
GHZ state of four superconducting qubits embedded in four
different resonators \textrm{can be created with this scheme}. In
principle, \textrm{the proposed method} can be extended to create
an entangled coherent state of $n$ resonators and to prepare a
Greenberger-Horne-Zeilinger (GHZ) state of $n$ qubits
distributed over $n$ cavities in a quantum
network. In addition, it is noted that four resonators coupled by
a coupler qubit may be used as a basic circuit block to build a
two-dimensional quantum network, which is useful for scalable
quantum information processing.
\end{abstract}

\maketitle
\date{\today}

\begin{center}
\textbf{I. INTRODUCTION}
\end{center}

Recent progress in circuit cavity QED, in which superconducting qubits play
the role of atoms in atom cavity QED, makes it stand out among the most
promising candidates for implementing quantum information processing (QIP)
[1]. Superconducting qubits, such as charge, flux, and phase qubits, and
microwave resonators (a.k.a. cavities) can be fabricated using modern
integrated circuit technology, their properties can be characterized and
adjusted in situ, they have relatively long decoherence times [2], and
various single and multiple qubits operations with state readout have been
demonstrated [3-7]. In particular, it has been demonstrated that a
superconducting resonator provides a quantized cavity field which can
mediate long-range and fast interaction between distant superconducting
qubits [8-10]. Theoretically, it was predicted earlier that the strong
coupling limit can readily be realized with superconducting charge qubits
[11] or flux qubits [12]. Moreover, the strong coupling limit between the
cavity field and superconducting qubits has been experimentally demonstrated
[13,14]. All of these theoretical and experimental progresses make circuit
cavity QED very attractive for QIP.

During the past decade, many theoretical proposals have been presented for
the preparation of Fock states, coherent states, squeezed states, the
Sch\"ordinger Cat state, and an arbitrary superposition of Fock states of a
single superconducting resonator [15-17]. Also, experimental creation of a
Fock state and a superposition of Fock states of a single superconducting
resonator using a superconducting qubit has been reported [18,19]. On the
other hand, a large number of theoretical proposals have been presented for
implementing quantum logical gates and generating quantum entanglement with
two or more superconducting qubits placed in a cavity or coupled by a
resonator (usually in the form of coplanar transmission line)
[8,11,12,20-24]. Moreover, experimental demonstration of two-qubit gates and
experimental preparation of three-qubit entanglement have been reported with
superconducting qubits in a cavity [9,25,26]. However, realistic QIP will
most likely need a large number of qubits and placing all of them in a
single cavity quickly runs into many fundamental and practical problems such
as the increase of cavity decay rate and decrease of qubit-cavity coupling
strength.

Therefore, future QIP most likely will require quantum networks consisting
of a large number of cavities each hosting and coupled to multiple qubits.
In this type of architecture transfer and exchange of quantum information
will not only occur among qubits in the same cavity but also between
different cavities. Hence, attention must be paid to the preparation of
quantum states of two or more superconducting resonators (hereafter we use
the term cavity and resonator interchangeably), preparation of quantum
states of superconducting qubits located in different cavities, and
implementation of quantum logic gates on superconducting qubits distributed
over different resonators in a network. All of these ingredients are
essential to realizing large-scale quantum information processing based on
circuit QED. Recently, a theoretical proposal for the manipulation and
generation of nonclassical microwave field states as well as the creation of
controlled multipartite entanglement with two resonators coupled by a
superconducting qubit has been presented [27], and a theoretical method for
synthesizing an arbitrary quantum state of two superconducting resonators
using a tunable superconducting qubit has been proposed [28]. Moreover,
experimental demonstration of the creation of $N$-photon NOON states
(entangled states $\left| N0\right\rangle +\left| 0N\right\rangle $) in two
superconducting microwave resonators by using a superconducting phase qubit
coupled to two resonators [29], and experimentally shuffling one- and
two-photon Fock states between three resonators interconnected by two
superconducting phase qubits have been reported recently [30]. These works
opened a new avenue for building one-dimensional linear quantum networks of
resonators and qubits.

On the other hand, entanglement between the atomic states and the coherent
states of a single-mode cavity was earlier demonstrated in experiments [31].
However, how to create an entangled coherent state between two or more
resonators, based on cavity QED, has not been reported yet. As is well
known, entangled coherent states are important in quantum information
processing and communication. For instances, they can be used to construct
quantum gates [32] (using coherent states as the logical qubits [33]),
perform teleportation [34], build quantum repeaters [35], implement quantum
key distribution [36], and entangle distant atoms in a network [37,38].
Moveover, it was first showed [39] that entangled coherent states can be
used to test violation of Bell inequalities.

In this paper, we propose a way for generating entangled coherent states of
four resonators using one three-level superconducting qubit as the
inter-cavity coupler. This proposal operates essentially by bringing the
transition between the two higher energy levels of the coupler qubit
dispersively coupled to the resonator modes. In addition, we will show how
to create a Greenberger-Horne-Zeilinger (GHZ) state of four superconducting
qubits located in four different resonators using the coupler qubit. The GHZ
states are multiqubit entangled states of the form $\left|
00...0\right\rangle \pm \left| 11...1\right\rangle ,$ which are useful in
quantum information processing [40] and communication [41].

Our proposal has the advantages: (i) Only one tunable superconducting qubit
is needed; (ii) The operation procedure and the operation time are both
independent of the number of resonators as well as the number of qubits in
the cavities; (iii) No adjustment of the resonator mode frequencies is
required during the entire operation; and (iv) The proposed method can in
principle be applied to create entangled coherent states of $n$ resonators
and to prepare a GHZ state of $n$ qubits distributed over $n$ cavities in a
quantum network, for which the operational steps and the operation time do
not increase as $n$ becomes larger.

This proposal is quite general, which can be applied to other types of
physical qubit systems with three levels, such as quantum dots and NV
centers coupled to \textrm{cavities.} The present work is of interest
because it provides a way to generate entangled coherent states of multiple
cavities and create a GHZ entangled state of qubits distributed over
multiple cavities, which are important in quantum information processing and
quantum communication. Finally, it is interesting to note that the four
resonators coupled by a coupler qubit can be used as a basic circuit block
to build a two-dimensional quantum network, which may be useful for scalable
quantum information processing.

This paper is organized as follows. In Sec.~II, we review some basic theory
of a coupler qubit interacting with four or three resonators. In Sec. III,
we discuss how to create four-resonator entangled coherent states. In
Sec.~IV, we show a way to generate a GHZ entangled state of qubits embedded
in four cavities without measurement. In Sec.~V, we give a discussion on the
possibility of using the four resonators coupled by a coupler qubit to build
a two-dimensional quantum network. In Sec.~VI, we give a brief discussion of
the experimental issues and possible experimental implementation. A
concluding summary is given in Sec.~VII.

\begin{figure}[tbp]
\begin{center}
\includegraphics[bb=40 238 538 549, width=10.5 cm, clip]{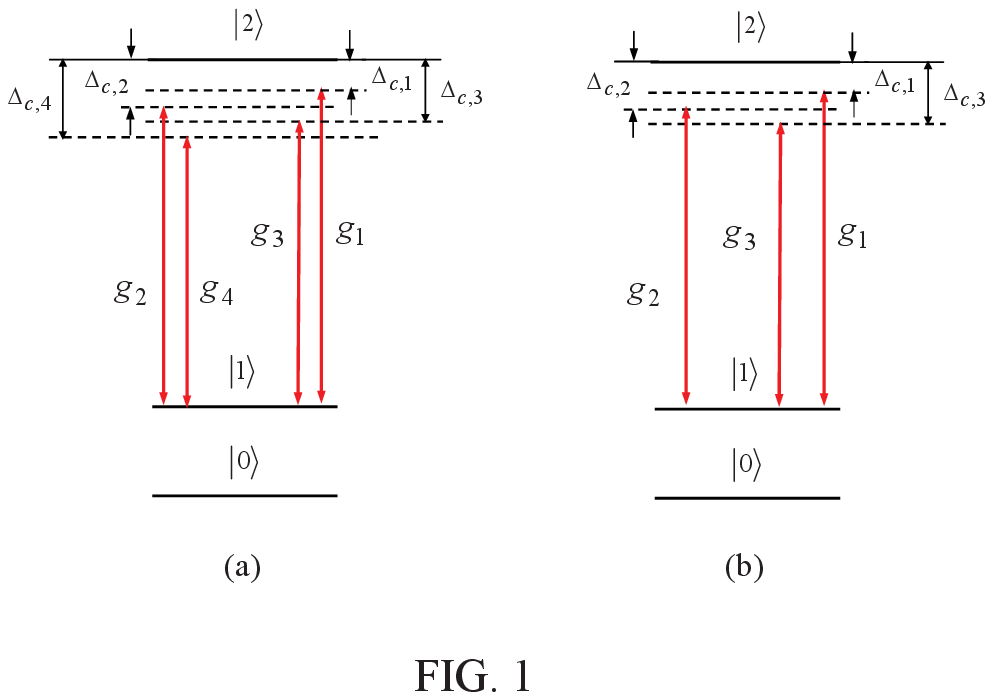} %
\vspace*{-0.08in}
\end{center}
\caption{(Color online) (a) Illustration of four resonators each
dispersively coupled with the $\left| 1\right\rangle \leftrightarrow \left|
2\right\rangle $ transition of qubit $A$. Here, $\Delta_{c,i}$ is the large
detuning between the $\left| 1\right\rangle \leftrightarrow \left|
2\right\rangle $ transition frequency of qubit $A$ and the frequency $%
\omega_{c,i}$ of resonator i, which satisfies $\Delta_{c,i}\gg g_i$ ($%
i=1,2,3,4$). (b) Illustration of three resonators each dispersively coupled
with the $\left| 1\right\rangle \leftrightarrow \left| 2\right\rangle $
transition of qubit $A$, with $\Delta_{c,i}\gg g_i$ ($i=1,2,3$). For
simplicity, we here consider the case that the $\left| 0\right\rangle
\leftrightarrow \left| 1\right\rangle $ level spacing is smaller than the $%
\left| 1\right\rangle \leftrightarrow \left| 2\right\rangle $ level spacing.
This type of level structure is available in superconducting charge qubits
or flux qubits [24]. Alternatively, the $\left| 0\right\rangle
\leftrightarrow \left| 1\right\rangle $ level spacing can be larger than the
$\left| 1\right\rangle \leftrightarrow \left| 2\right\rangle $ level
spacing, which applies to superconducting phase qubits [24].}
\label{fig:1}
\end{figure}

\begin{center}
II\textbf{. BASIC THEORY}
\end{center}

Consider a three-level superconducting qubit $A$, with states $|0\rangle ,$ $%
|1\rangle ,$ and $|2\rangle $, coupled to four resonators $1,$ $2,$ $3,$ and
$4$ as shown in Fig. 1(a) or three resonators $1,2,$ and $3$ as depicted in
Fig. 1(b). Suppose that the relevant mode frequency of each resonator is
coupled to the $\left| 1\right\rangle \leftrightarrow \left| 2\right\rangle $
transition while decoupled from transitions between other levels of the
qubit (Fig.~1). The Hamiltonian for the whole system is given by (assuming $%
\hbar =1$ for simplicity)
\begin{equation}
H=\sum_{i=1}^m\omega _{c,i}a_i^{+}a_i+\frac{\omega _0}2S_z+\sum_{i=1}^mg_i%
\left( a_iS_{+}+a_i^{+}S_{-}\right) ,
\end{equation}
where $m=4$ corresponds to qubit $A$ coupled to the four resonators 1, 2, 3,
and 4 while $m=3$ corresponds to qubit $A$ coupled to the three resonators
1, 2, and 3; $S_{+}=\left| 2\right\rangle \left\langle 1\right| ,$ $%
S_{-}=\left| 1\right\rangle \left\langle 2\right| $, $S_z=\left|
2\right\rangle \left\langle 2\right| -\left| 1\right\rangle \left\langle
1\right| $; $a_i$ ($a_i^{\dagger }$) is the photon annihilation (creation)
operator of resonator $i$ with frequency $\omega _{c,i};$ $\omega _0$ is the
transition frequency between the two levels $|1\rangle $ and $|2\rangle $ of
qubit $A;$ and $g_i$ is the coupling constant between the resonator $i$ and
the $\left| 1\right\rangle \leftrightarrow \left| 2\right\rangle $
transition of qubit $A$. In the interaction picture, the Hamiltonian (1)
becomes

\begin{equation}
H_{\text{\textrm{I}}}=\sum_{i=1}^mg_i\left( e^{i\Delta
_{c,i}t}a_iS_{+}+e^{-i\Delta _{c,i}t}a_i^{+}S_{-}\right) ,
\end{equation}
where $\Delta _{c,i}=\omega _0-\omega _{c,i}$ is the detuning between the $%
\left| 1\right\rangle \leftrightarrow \left| 2\right\rangle $ transition
frequency $\omega _0$ of qubit $A$ and the $i$th resonator frequency $\omega
_{c,i}$ Suppose that (i) the $\left| 1\right\rangle \leftrightarrow \left|
2\right\rangle $ transition of qubit $A$ is dispersively coupled with the
resonator $i$ (i.e., $\Delta _{c,i}\gg g_i$) (Fig.~1); and (ii) $\Delta
_{c,i+1}-\Delta _{c,i}$ is on the same order of magnitude as the coupling
constant $g_i,$ such that the indirect interaction between any two
resonators induced by qubit $A$ is negligible. Under these conditions, the
Hamiltonian (2) reduces to [42]
\begin{equation}
H_{\mathrm{eff}}=\sum_{i=1}^m\frac{g_i^2}{\Delta _{c,i}}\left(
a_ia_i^{+}\left| 2\right\rangle \left\langle 2\right| -a_i^{+}a_i\left|
1\right\rangle \left\langle 1\right| \right) .
\end{equation}
One can see that the Stark shift terms $\sum_{i=1}^mg_i^2a_ia_i^{+}\left|
2\right\rangle \left\langle 2\right| /\Delta _{c,i}$ involved in the
Hamiltonian (2) do not affect the state $\left| 1\right\rangle $ of qubit $A$
during the evolution.

Based on the Hamiltonian (3), it is easy to see that if the resonator $i$ is
initially in a coherent state $\left| \alpha _i\right\rangle $, the time
evolution of the state $\left| 1\right\rangle _A\left| \alpha
_i\right\rangle $ of the system composed of qubit $A$ and the resonator $i$
is then described by
\begin{equation}
\left| 1\right\rangle _A\left| \alpha _i\right\rangle \rightarrow \left|
1\right\rangle _A\left| \alpha _i\exp (ig_i^2t/\Delta _{c,i})\right\rangle ,
\end{equation}
which leads to the coherent state \textrm{of the }$i$\textrm{-th cavity
evolve from} $\left| \alpha _i\right\rangle $ to $\left| -\alpha
_i\right\rangle $ when $g_i^2t/\Delta _{c,i}=\pi .$ The state $\left|
0\right\rangle _A\left| \alpha _i\right\rangle $ does not change under the
Hamiltonian (3). The result (4) presented here will be employed for creation
of four-resonator entangled coherent states as discussed in next section.

In addition, based on the Hamiltonian (3), it is easy to find that if the
resonator $i$ is initially in a single-photon state $\left| 1\right\rangle
_{c,i}$, the time evolution of the state $\left| 1\right\rangle _A\left|
1\right\rangle _{c,i}$ of the system composed of qubit $A$ and the resonator
$i$ is then given by
\begin{equation}
\left| 1\right\rangle _A\left| 1\right\rangle _{c,i}\rightarrow
e^{ig_i^2t/\Delta _{c,i}}\left| 1\right\rangle _A\left| 1\right\rangle
_{c,i},
\end{equation}
which introduces a phase flip to the state $\left| 1\right\rangle _A\left|
1\right\rangle _{c,i}$ when the evolution time $t$ satisfies $g_i^2t/\Delta
_{c,i}=\pi .$ Note that the states $\left| 0\right\rangle _A\left|
0\right\rangle _{c,i},$ $\left| 1\right\rangle _A\left| 0\right\rangle
_{c,i},$ and $\left| 0\right\rangle _A\left| 1\right\rangle _{c,i}$ remain
unchanged under the Hamiltonian (3). This result (5) will be employed for
generation of a GHZ state of four qubits distributed over four different
cavities.

It should be mentioned here that during the following entanglement
preparation, the level $\left| 0\right\rangle $ of the coupler qubit $A$ is
not affected by the mode of each resonator. To meet this condition, one can
choose qubit $A$ for which the transition between the two lowest levels $%
\left| 0\right\rangle $ and $\left| 1\right\rangle $ is forbidden due to the
optical selection rules [43], weak via increasing the potential barrier
between the two lowest levels [2,44-46], or highly detuned (decoupled) from
the cavity mode of each resonator, which can be achieved by adjusting the
level spacings of qubit $A.$ Note that for superconducting qubits the level
spacings can be readily adjusted by varying external control parameters
[2,45,47].

\begin{figure}[tbp]
\includegraphics[bb=84 399 526 647, width=10.5 cm, clip]{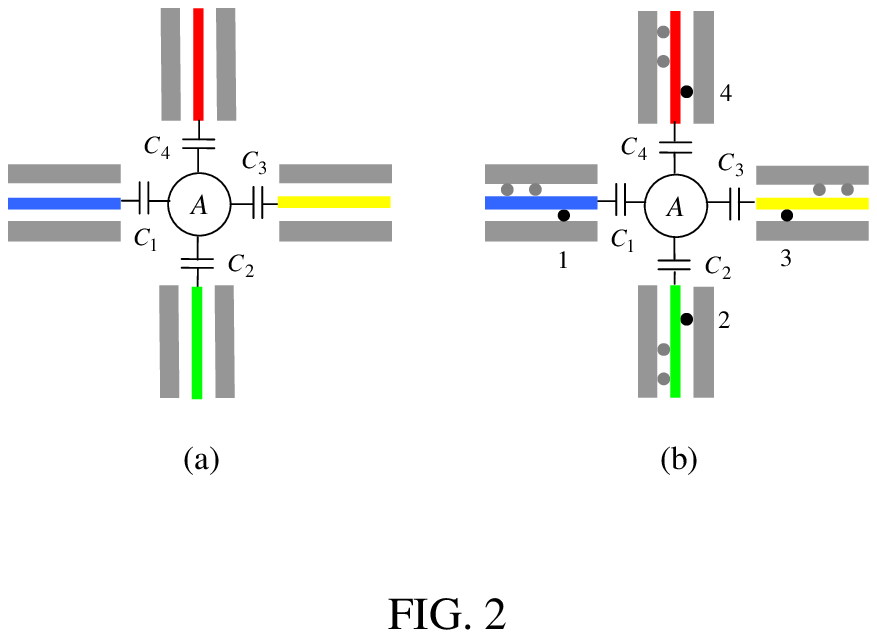} %
\vspace*{-0.08in}
\caption{(Color online) (a) and (b) Diagram of a superconducting qubit $A$
(a circle at the center) coupled capacitively to four one-dimensional
coplanar waveguide resonators through $C_{1},C_{2},C_{3},C_{4}$,
respectively. In (b), a black or grey dot in each resonator represents a
qubit. The four black-dot qubits ($1,2,3,4$) are first prepared in a GHZ
state, which can further be entangled with all other qubits (grey dots). For
clarity, only three qubits in each cavity are shown.}
\label{fig:2}
\end{figure}

\begin{center}
\textbf{III. CREATION OF FOUR-RESONATOR ENTANGLED COHERENT STATES}
\end{center}

In this section, we will show how to generate an entangled coherent state of
four resonators, give a discussion of the fidelity of the operations, and
then address issues which are relevant to this topic.

\begin{center}
\textbf{A. Generation of four-resonator entangled coherent states }
\end{center}

Consider a system composed of four resonators and a superconducting qubit $A$
[Fig.~2(a)]. The qubit $A$ has three levels shown in Fig.~1. Initially, the
qubit $A$ is decoupled from all resonators [Fig. 3(a)], which can be
realized by prior adjustment of the qubit level spacings [2,45,47]. The
qubit $A$ is initially prepared in the state $(\left| 0\right\rangle
_A+\left| 1\right\rangle _A)/\sqrt{2}$ and each resonator is initially
prepared in a coherent state [15,19], \textit{i.e}., $\left| \alpha
_i\right\rangle $ for resonator $i$ ($i=1,2,3,4$). To prepare the four
resonators in an entangled coherent state, we now perform the following
operations:

\begin{figure}[tbp]
\begin{center}
\includegraphics[bb=82 240 533 564, width=10.5 cm, clip]{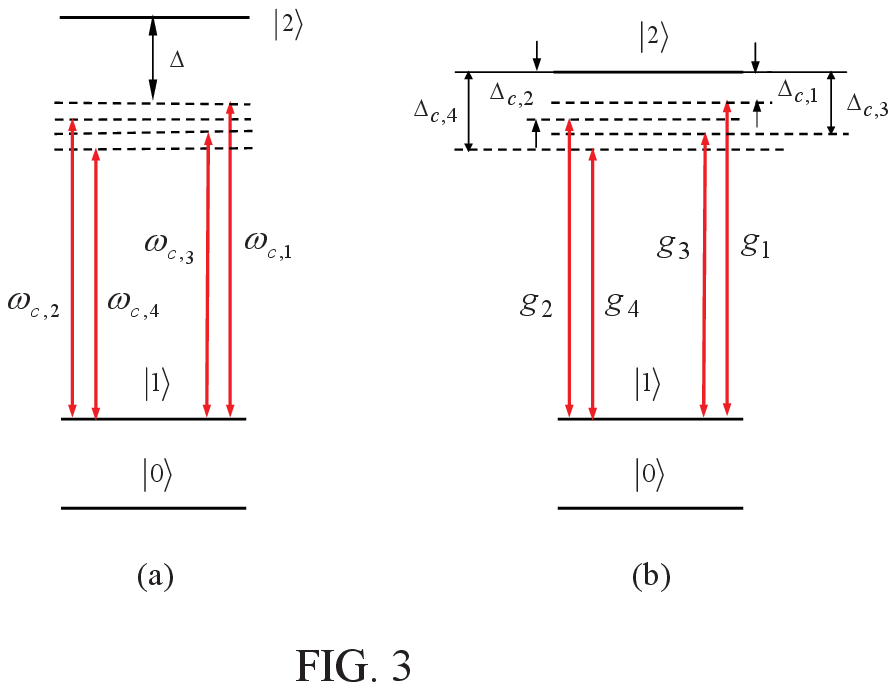} %
\vspace*{-0.08in}
\end{center}
\caption{(Color online) (a) Illustration of qubit $A$ decoupled from four
cavities or resonators. Here, $\Delta$ is the large detuning between the $%
\left| 1\right\rangle \leftrightarrow \left| 2\right\rangle $ transition
frequency of qubit $A$ and the frequency $\omega_{c,1}$ of resonator 1,
which represents that the $\left| 1\right\rangle \leftrightarrow \left|
2\right\rangle $ transition of qubit $A$ is far-off resonant with (decoupled
from) resonator 1. Since the frequencies $\omega_{c,1},\omega_{c,2},%
\omega_{c,3},$ and $\omega_{c,4}$ of four resonators 1, 2, 3, and 4 satisfy $%
\omega _{c,1}>\omega _{c,2}>\omega _{c,3}>\omega _{c,4}$, the $\left|
1\right\rangle \leftrightarrow \left| 2\right\rangle $ transition of qubit $%
A $ is also far-off resonant with (decoupled from) the other three
resonators 2, 3, and 4. (b) Illustration of four resonators each
dispersively coupled with the $\left| 1\right\rangle \leftrightarrow \left|
2\right\rangle $ transition of qubit $A$. Here, $\Delta_{c,i}$ is the large
detuning between the $\left| 1\right\rangle \leftrightarrow \left|
2\right\rangle $ transition frequency of qubit $A$ and the frequency $%
\omega_{c,i}$ of resonator i, which satisfies $\Delta_{c,i}\gg g_i$ ($%
i=1,2,3,4$).}
\label{fig:3}
\end{figure}

Step (i): Adjust the level spacings of the qubit $A$ such that the field
mode for each resonator is dispersively coupled to the $\left|
1\right\rangle \leftrightarrow \left| 2\right\rangle $ transition (i.e., $%
\Delta _{c,i}=\omega _{21}-\omega _{c,i}\gg g_i$ for resonator $i$) while
far-off resonant with (decoupled from) the transition between other levels
of the qubit $A$ [Fig.~3(b)]. After an interaction time $\tau $, the initial
state $(\left| 0\right\rangle _A+\left| 1\right\rangle
_A)\prod_{i=1}^4\left| \alpha _i\right\rangle $ of the whole system changes
to (here and below a normalized factor is omitted for simplicity)
\begin{equation}
\left| 0\right\rangle _A\prod_{i=1}^4\left| \alpha _i\right\rangle +\left|
1\right\rangle _A\prod_{i=1}^4\left| \alpha _i\exp (ig_i^2t/\Delta
_{c,i})\right\rangle .
\end{equation}
Both of the resonators and qubits can be fabricated to have appropriate
resonator frequencies and qubit-cavity coupling strengths, such that $\frac{%
g_1^2}{\Delta _{c,1}}=\frac{g_2^2}{\Delta _{c,2}}=\frac{g_3^2}{\Delta _{c,3}}%
=\frac{g_4^2}{\Delta _{c,4}}.$ Note that tunable qubit-cavity coupling
strength has been proposed and demonstrated experimentally [48-50]. For $%
g_i^2\tau /\Delta _{c,i}=\pi $ ($i=1,2,3,4$)$,$ the system then evolves to
\begin{equation}
\left| 0\right\rangle _A\prod_{i=1}^4\left| \alpha _i\right\rangle +\left|
1\right\rangle _A\prod_{i=1}^4\left| -\alpha _i\right\rangle ,
\end{equation}
according to Eq. (6). Here, $\left\langle \alpha _i\right| \left. -\alpha
_i\right\rangle =\exp \left( -2\left| \alpha _i\right| ^2\right) \approx 0$
when $\alpha _i$ is large enough.

Step (ii): Adjust the level spacings of the qubit $A$ to the original
configuration such that it is decoupled (\textit{i.e}., far off-resonance)
from all resonators [Fig. 3(a)]. We then apply a classical $\pi /2$-pulse
(resonant with the $\left| 0\right\rangle \leftrightarrow \left|
1\right\rangle $ transition of the qubit $A$) to transform the qubit state $%
\left| 0\right\rangle _A$ to $\left| 0\right\rangle _A+\left| 1\right\rangle
_A$ and $\left| 1\right\rangle _A$ to $-\left| 0\right\rangle _A+\left|
1\right\rangle _A$. Thus, the state (7) becomes

\begin{equation}
\left| 0\right\rangle _A\left( \prod_{i=1}^4\left| \alpha _i\right\rangle
-\prod_{i=1}^4\left| -\alpha _i\right\rangle \right) +\left| 1\right\rangle
_A\left( \prod_{i=1}^4\left| \alpha _i\right\rangle +\prod_{i=1}^4\left|
-\alpha _i\right\rangle \right) .
\end{equation}
We now perform a measurement on the states of the qubit $A$ in the \{$\left|
0\right\rangle ,$ $\left| 1\right\rangle $\} basis. If the qubit $A$ is
found in the state $\left| 0\right\rangle ,$ it can be seen from Eq.~(8)
that the four resonators must be in the following entangled coherent state
\begin{equation}
\mathcal{N}_{-}(\left| \alpha _1\right\rangle \left| \alpha _2\right\rangle
\left| \alpha _3\right\rangle \left| \alpha _4\right\rangle -\left| -\alpha
_1\right\rangle \left| -\alpha _2\right\rangle \left| -\alpha
_3\right\rangle \left| -\alpha _4\right\rangle ),
\end{equation}
Similarly, if the qubit is found in the state $\left| 1\right\rangle ,$ then
the four resonators must be in the following entangled coherent state
\begin{equation}
\mathcal{N}_{+}(\left| \alpha _1\right\rangle \left| \alpha _2\right\rangle
\left| \alpha _3\right\rangle \left| \alpha _4\right\rangle +\left| -\alpha
_1\right\rangle \left| -\alpha _2\right\rangle \left| -\alpha
_3\right\rangle \left| -\alpha _4\right\rangle ),
\end{equation}
where $\mathcal{N}_{\mp }$ are the normalization factors.

We should point out that since the level spacing between the two levels $%
\left| 1\right\rangle $ and $\left| 2\right\rangle $ of qubit $A$ in Fig.
3(a) is set to be greater than that in Fig. 3(b), qubit $A$ remains
off-resonant with any of the four resonators during tuning the level
structure of qubit $A$ from Fig. 3(a) to Fig. 3(b).

It is straightforward to show that by using a superconducting qubit coupled
to $n$ resonators ($1,2,...,n$) initially in the state $\prod_{i=1}^n\left|
\alpha _i\right\rangle $, the $n$-resonator entangled coherent state $%
\prod_{i=1}^n\left| \alpha _i\right\rangle -\prod_{i=1}^n\left| -\alpha
_i\right\rangle $ or $\prod_{i=1}^n\left| \alpha _i\right\rangle
+\prod_{i=1}^n\left| -\alpha _i\right\rangle $ can be prepared by using the
same procedure given above.

\begin{center}
\textbf{B. Fidelity}
\end{center}

Let us now give a discussion of the fidelity of the operations. Since only
the qubit-pulse resonant interaction is used in step (ii), this step can be
completed within a very short time (e.g., by increasing the pulse Rabi
frequency), such that the dissipation of the qubit and the cavities is
negligibly small. In this case, the dissipation of the system would appear
in the operation of step (i) because of the qubit-cavity dispersive
interaction. During the operation of step (i), the dynamics of the lossy
system is determined by
\begin{eqnarray}
\frac{d\rho }{dt} &=&-i\left[ H_I,\rho \right] +\sum_{i=1}^4\kappa _i%
\mathcal{L}\left[ a_i\right] +\left\{ \gamma _\varphi \left( S_z\rho
S_z-\rho \right) +\gamma \mathcal{L}\left[ S_{-}\right] \right\}  \nonumber
\\
&&+\left\{ \gamma _\varphi ^{\prime }\left( S_z^{\prime }\rho S_z^{\prime
}-\rho \right) +\gamma ^{\prime }\mathcal{L}\left[ S_{-}^{\prime }\right]
\right\} +\left\{ \gamma _\varphi ^{\prime \prime }\left( S_z^{\prime \prime
}\rho S_z^{\prime \prime }-\rho \right) +\gamma ^{\prime \prime }\mathcal{L}%
\left[ S_{-}^{\prime \prime }\right] \right\} ,
\end{eqnarray}
where $H_I$ is the Hamiltonian (2)$,$ $\mathcal{L}\left[ a_i\right] =a_i\rho
a_i^{+}-a_i^{+}a_i\rho /2-\rho a_i^{+}a_i/2,$ $\mathcal{L}\left[
S_{-}\right] =S_{-}\rho S_{+}-S_{+}S_{-}\rho /2-\rho S_{+}S_{-}/2,$ $%
\mathcal{L}\left[ S_{-}^{\prime }\right] =S_{-}^{\prime }\rho S_{-}^{\prime
}-S_{+}^{\prime }S_{-}^{\prime }\rho /2-\rho S_{+}^{\prime }S_{-}^{\prime
}/2,$ and $\mathcal{L}\left[ S_{-}^{\prime \prime }\right] =S_{-}^{\prime
\prime }\rho S_{+}^{\prime \prime }-S_{+}^{\prime \prime }S_{-}^{\prime
\prime }\rho /2-\rho S_{+}^{\prime \prime }S_{-}^{\prime \prime }/2$ (with $%
S_z^{\prime }=\left| 2\right\rangle \left\langle 2\right| -\left|
0\right\rangle \left\langle 0\right| $, $S_z^{\prime \prime }=\left|
1\right\rangle \left\langle 1\right| -\left| 0\right\rangle \left\langle
0\right| $, $S_{-}^{\prime }=\left| 0\right\rangle \left\langle 2\right| ,$
and $S_{-}^{\prime \prime }=\left| 0\right\rangle \left\langle 1\right| $ ).
In addition, $\kappa _i$ is the decay rate of the mode of cavity $i,$ $%
\gamma _\varphi $ and $\gamma $ are the dephasing rate and the energy
relaxation rate of the level $\left| 2\right\rangle $ of qubit $A$ for the
decay path $\left| 2\right\rangle \rightarrow \left| 1\right\rangle ,$ $%
\gamma _\varphi ^{\prime }$ and $\gamma ^{\prime }$ are the dephasing rate
and the energy relaxation rate of the level $\left| 2\right\rangle $ of
qubit $A$ for the decay path $\left| 2\right\rangle \rightarrow \left|
0\right\rangle ,$ and $\gamma _\varphi ^{\prime \prime }$ and $\gamma
^{\prime \prime }$ are the dephasing rate and the energy relaxation rate of
the level $\left| 1\right\rangle $ of qubit $A$ for the decay path $\left|
1\right\rangle \rightarrow \left| 0\right\rangle $, respectively. The
fidelity of the operations is given by
\begin{equation}
\mathcal{F}=\left\langle \psi _{id}\right| \widetilde{\rho }\left| \psi
_{id}\right\rangle ,
\end{equation}
where $\left| \psi _{id}\right\rangle $ is the state (8) of the whole system
after the above operations, in the ideal case without considering the
dissipation of the system during the entire operation; and $\widetilde{\rho }
$ is the final density operator of the whole system when the operations are
performed in a real situation.

A coherent state $\left| \alpha _i\right\rangle $ can be expressed as $%
\left| \alpha _i\right\rangle =\exp \left[ -\left| \alpha _i\right|
^2/2\right] \sum_{n=0}^\infty \frac{\alpha _i^n}{\sqrt{n!}}\left|
n\right\rangle $ in a Fock-state basis. In our numerical calculation, we
consider the first $m$ terms in the expansions of $\left| \alpha
_i\right\rangle $ and $\left| -\alpha _i\right\rangle ,$ i.e.,
\begin{eqnarray}
\left| \alpha _i\right\rangle &\approx &\exp \left[ -\left| \alpha _i\right|
^2/2\right] \sum_{n=0}^m\frac{\alpha _i^n}{\sqrt{n!}}\left| n\right\rangle ,
\nonumber \\
\left| -\alpha _i\right\rangle &\approx &\exp \left[ -\left| \alpha
_i\right| ^2/2\right] \sum_{n=0}^m\frac{\left( -\alpha _i\right) ^n}{\sqrt{n!%
}}\left| n\right\rangle .
\end{eqnarray}
Under this consideration, the expression of the fidelity above is modified
as
\begin{equation}
\mathcal{F}=\frac{\left\langle \psi _{id}\right| \widetilde{\rho }\left|
\psi _{id}\right\rangle }{\left| \left\langle \psi _{id}\right| \left. \psi
_{id}\right\rangle \right| ^2},
\end{equation}
where $\left| \psi _{id}\right\rangle $ is the state (8) in which the
coherence states $\left| \alpha _i\right\rangle $ and $\left| -\alpha
_i\right\rangle $ are now replaced by the states given in Eq.~(13), and the
denominator $\left| \left\langle \psi _{id}\right| \left. \psi
_{id}\right\rangle \right| ^2$ arises from the normalization of the state $%
\left| \psi _{id}\right\rangle .$ For simplicity, we consider $\alpha
_1=\alpha _2=\alpha _3=\alpha _4=\alpha $ in our numerical calculation.

\begin{figure}[tbp]
\begin{center}
\includegraphics[bb=0 0 471 305, width=9.0 cm, clip]{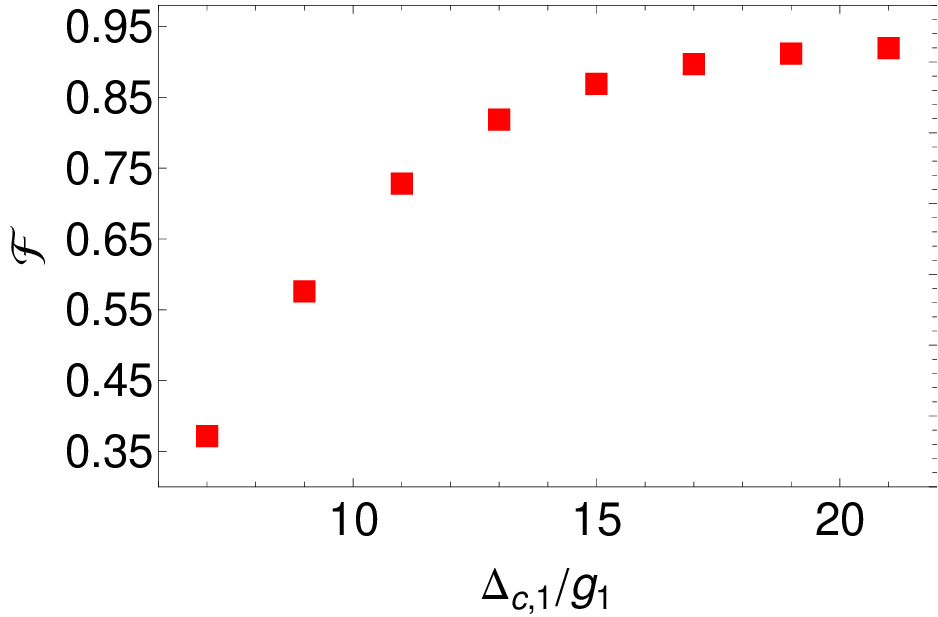} %
\vspace*{-0.08in}
\end{center}
\caption{(Color online) Fidelity versus $\Delta_{c,1}/g_1$. The parameters
used in the numerical calculation are $\gamma _\varphi ^{-1}=(\gamma
_\varphi ^{\prime })^{-1}=(\gamma _\varphi ^{\prime \prime })^{-1}=5$ $\mu $%
s, $\gamma ^{-1}=25$ $\mu $s, $(\gamma ^{\prime })^{-1}=200$ $\mu $s, $%
(\gamma ^{\prime \prime })^{-1}=50$ $\mu $s, $\kappa _1^{-1}=\kappa
_2^{-1}=\kappa _3^{-1}=\kappa _4^{-1}=20$ $\mu $s, $s=0.5,$ and $g_1/2\pi
=75 $ MHz.}
\label{fig:4}
\end{figure}

By defining $\Delta _{c,4}-\Delta _{c,3}=\Delta _{c,3}-\Delta _{c,2}=\Delta
_{c,2}-\Delta _{c,1}=s\Delta _{c,1},$ we have $\omega _{c,2}=\omega
_{c,1}-s\Delta _{c,1},$ $\omega _{c,3}=\omega _{c,1}-2s\Delta _{c,1},$ and $%
\omega _{c,4}=\omega _{c,1}-3s\Delta _{c,1}.$ According to $g_1^2/\Delta
_{c,1}=g_2^2/\Delta _{c,2}=g_3^2/\Delta _{c,3}=g_4^2/\Delta _{c,4},$ we have
$g_2=\sqrt{1+s}g_1,$ $g_3=\sqrt{1+2s}g_1,$ and $g_4=\sqrt{1+3s}g_1.$ For the
choice of $\gamma _\varphi ^{-1}=(\gamma _\varphi ^{\prime })^{-1}=(\gamma
_\varphi ^{\prime \prime })^{-1}=5$ $\mu $s, $\gamma ^{-1}=25$ $\mu $s, $%
(\gamma ^{\prime })^{-1}=200$ $\mu $s, $(\gamma ^{\prime \prime })^{-1}=50$ $%
\mu $s, $\kappa _1^{-1}=\kappa _2^{-1}=\kappa _3^{-1}=\kappa _4^{-1}=20$ $%
\mu $s, $s=0.5,$ and $g_1/2\pi =75$ MHz, the fidelity versus the parameter $%
\Delta _{c,1}/g_1$ is shown in Fig.~4 where only eight points are plotted
and each point is based on the numerical calculation for $\alpha =1.1$ and $%
m=3.$ From Fig.~4, it can be seen that a high fidelity $\sim 93\%$ can be
achieved when $\Delta _{c,1}/g_1=20.$ For $s=0.5$ here, we have $g_2/2\pi
\sim 92$ MHz, $g_3\sim 106$ MHz, and $g_4\sim 119$ MHz. Note that a
qubit-cavity coupling constant $\sim 220$ MHz can be reached for a
superconducting qubit coupled to a one-dimensional standing-wave CPW
(coplanar waveguide) transmission line resonator [26], and that $T_1$ and $%
T_2$ can be made to be on the order of $10-100$ $\mu $s for the state of art
superconducting qubits at this time [51]. Without loss of generality, assume
that the $\left| 1\right\rangle \leftrightarrow \left| 2\right\rangle $
transition frequency of qubit $A$ is $\nu _0\sim 10$ GHz, and thus the
frequency of cavity $1,$ the frequency of cavity $2,$ the frequency of
cavity $3$ and the frequency of cavity $4$ are $\nu _{c,1}\sim 8.5$ GHz, $%
\nu _{c,2}\sim 7.75$ GHz, $\nu _{c,3}\sim 7$ GHz and $\nu _{c,4}\sim 6.25$
GHz, respectively [25]. For the cavity frequencies chosen here and for the $%
\kappa _1^{-1},\kappa _2^{-1},\kappa _3^{-1},$ $\kappa _4^{-1}$ used in the
numerical calculation, the required quality factors for cavities $1,2,3$ and
$4$ are $Q_1\sim 1.0\times 10^6,$ $Q_2\sim 9.7\times 10^5,$ $Q_3\sim
8.8\times 10^5,$ and $Q_4\sim 7.8\times 10^5$, respectively. Note that
superconducting CPW transmission line resonators with a loaded quality
factor $Q\sim 10^6$ have been experimentally demonstrated [52,53], and
planar superconducting resonators with internal quality factors above one
million ($Q>10^6$) has also been reported recently [54]. Our analysis given
here demonstrates that preparation of an entangled coherent state of four
cavities is possible within the present circuit cavity QED technique.

\begin{center}
\textbf{C. Discussion}
\end{center}

Note that the level $\left| 1\right\rangle $ of qubit $A$ has longer energy
relaxation time and dephasing time than the level $\left| 2\right\rangle .$
Thus, we focus on the level $\left| 2\right\rangle $ in the following.
According to [24], the energy relaxation of the level $\left| 2\right\rangle
$ of qubit $A$ can be enhanced via dressed dephasing of qubit $A$ by each
resonator. For simplicity, let us consider resonator $i.$ The effective
relaxation rate $\Gamma _e$ of the level $\left| 2\right\rangle $ of qubit $%
A,$ induced due to the dressed dephasing of qubit $A$ by the photons of
resonator $i$, is given by [55]
\begin{equation}
\gamma _e=\gamma \left( 1-\frac{2\overline{n}_i+1}{4n_{\text{crit,}i}}%
\right) +\gamma _{k,i}+\gamma _{\Delta ,i}\overline{n}_i,
\end{equation}
where $\gamma $ is the pure energy relaxation rate of the level $\left|
2\right\rangle $ of qubit $A$ caused by noise environment, $\gamma _{k,i}$
is the Purcell decay rate associated with resonator $i$, $\gamma _{\Delta ,i}
$ is the measurement and dephasing-induced relaxation rate, $n_{\text{crit,}%
i}=\Delta _{c,i}^2/4g_i^2$ is the critical photon number for resonator $i,$
and $\overline{n}_i$ is the average photon number of resonator $i.$ One can
see from Eq. (21) that to avoid the enhancement of the energy relaxation of
the level $\left| 2\right\rangle $ (i.e., to obtain $\gamma _e\leq \gamma $)$%
,$ the following condition
\begin{equation}
\overline{n}_i\leq \frac{\gamma -4n_{\text{crit,}i}\gamma _{k,i}}{4n_{\text{%
crit,}i}\gamma _{\Delta ,i}-2\gamma }
\end{equation}
needs to be satisfied. The result (16) provides a limitation on the average
photon number of resonator $i$ ($i=1,2,3,4$).

As shown above, measurement on the states of qubit $A$ is needed during
preparation of the entangled coherent states of cavities. To the best of our
knowledge, all existing proposals for creating entangled coherent states
based on cavity QED require a measurement on the states of qubits [56].

In the introduction, we have given a discussion on the significance of
entangled coherent states in quantum information processing and
communication. Here, we would like to add a few lines regarding
advantages/disadvantages of a network of coherent states might have versus
Fock states. The advantages are: when compared with Fock states, (i)
coherent states are more easily prepared in experiments; and (ii) they are
more robust against decoherence caused by noise environment and thus can be
transmitted for a longer distance. The disadvantage is: both an entangled
coherent state and an entangled Fock state may suffer from strong
decoherence when the average photon number is large.

\begin{center}
\textbf{IV. ENTANGLING QUBITS EMBEDDED IN DIFFERENT CAVITIES}
\end{center}

\begin{figure}[tbp]
\begin{center}
\includegraphics[bb=18 236 578 604, width=12.5 cm, clip]{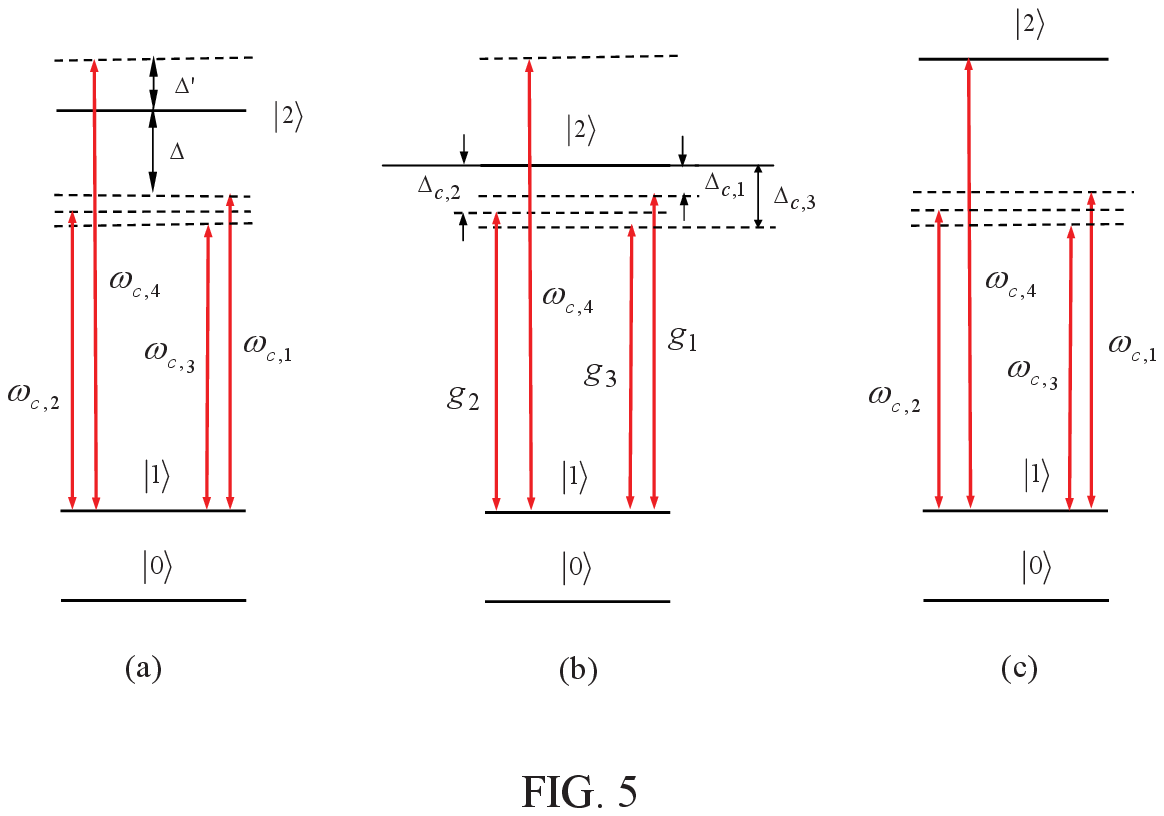} %
\vspace*{-0.08in}
\end{center}
\caption{(Color online) (a) Illustration of qubit $A$ decoupled from four
cavities or resonators. Here, $\Delta$ is the large detuning between the $%
\left| 1\right\rangle \leftrightarrow \left| 2\right\rangle $ transition
frequency of qubit $A$ and the frequency $\omega_{c,1}$ of resonator 1,
which represents that the $\left| 1\right\rangle \leftrightarrow \left|
2\right\rangle $ transition of qubit $A$ is far-off resonant with (decoupled
from) resonator 1. Since the frequencies $\omega_{c,1},\omega_{c,2}$, and $%
\omega_{c,3}$ of three resonators 1, 2, and 3 satisfy $\omega _{c,1}>\omega
_{c,2}>\omega _{c,3}$, the $\left| 1\right\rangle \leftrightarrow \left|
2\right\rangle $ transition of qubit $A $ is also far-off resonant with
(decoupled from) the other three resonators 2, 3, and 4. In addition, $%
\Delta^{\prime}$ is the large detuning between the $\left| 1\right\rangle
\leftrightarrow \left| 2\right\rangle $ transition frequency of qubit $A$
and the frequency $\omega_{c,4}$ of resonator 4, which indicates that the $%
\left| 1\right\rangle \leftrightarrow \left| 2\right\rangle $ transition of
qubit $A$ is far-off resonant with (decoupled from) resonator 4. (b)
Illustration of three resonators (1,2,3) each dispersively coupled with the $%
\left| 1\right\rangle \leftrightarrow \left| 2\right\rangle $ transition of
qubit $A$. Here, $\Delta_{c,i}$ is the large detuning between the $\left|
1\right\rangle \leftrightarrow \left| 2\right\rangle $ transition frequency
of qubit $A$ and the frequency $\omega_{c,i}$ of resonator i, which
satisfies $\Delta_{c,i}\gg g_i$ ($i=1,2,3$). When tuning the level spacings
of qubit $A$ from Fig.~5(a) to Fig.~5(b), the detuning $\Delta^{\prime}$
increases, thus qubit $A$ remains decoupled from resonator $4$. (c)
Illustration of resonator 4 resonantly coupled with the $\left|
1\right\rangle \leftrightarrow \left| 2\right\rangle $ transition of qubit $%
A $. When tuning the level spacings of qubit $A$ from Fig.~5(a) to
Fig.~5(c), the detuning $\Delta$ increases and thus the $\left|
1\right\rangle \leftrightarrow \left| 2\right\rangle $ transition of qubit $%
A $ remains decoupled from the three cavities 1, 2, and 3.}
\label{fig:5}
\end{figure}

In this section, we will show how to prepare a GHZ entangled state of four
qubits located at four different cavities. We then give a discussion of the
fidelity of the operations. Last, we discuss how to prepare multiple qubits
distributed over $n$ different cavities.

\begin{center}
\textbf{A. Preparation of GHZ states of four qubits in four cavities}
\end{center}

Consider a system composed of four cavities coupled by a three-level
superconducting qubit $A$ [Fig.~2(b)]. The qubit $A$ is initially decoupled
from the four cavities [Fig.~5(a)]. Each cavity hosts a two-level qubit $1,$
$2,$ $3,$ or $4,$ which is represented by a black dot [Fig.~2(b)]. The two
levels of each of qubits $1,2,3,$ and $4$ are labeled as $\left|
0\right\rangle $ (the ground state) and $\left| 1\right\rangle $ (the
excited state). The qubits ($1,2,3,4$) are initially decoupled from their
respective cavities. Qubit $A$ and qubits ($1,2,3$) are initially prepared
in the state $\left( \left| 0\right\rangle +\left| 1\right\rangle \right) /%
\sqrt{2},$ while qubit $4$ is initially in the state $\left| 0\right\rangle $%
. In addition, each cavity is initially in a vacuum state. The operations
for preparing the qubits ($1,2,3,4$) in a GHZ state are listed as follows:

Step (i): Adjust the level spacings of qubits ($1,2,3$) to bring the $\left|
0\right\rangle \leftrightarrow \left| 1\right\rangle $ transition of qubit $i
$ resonant with the cavity $i$ ($i=1,2,3$) for an interaction time $t_i=\pi
/(2g_{r,i}),$ such that the state $\left| 1\right\rangle _i\left|
0\right\rangle _{c,i}$ is transformed to $-i\left| 0\right\rangle _i\left|
1\right\rangle _{c,i}$ while the state $\left| 0\right\rangle _i\left|
0\right\rangle _{c,i}$ remains unchanged. Here, $g_{r,i}$ is the resonant
coupling constant of qubit $i$ with its own cavity $i$. After the operation
of this step, the initial state of the whole system changes to [57]
\begin{equation}
\prod_{i=1}^3\left[ \left| 0\right\rangle _i(\left| 0\right\rangle
_{c,i}-i\left| 1\right\rangle _{c,i})\right] \otimes \left| 0\right\rangle
_4\left| 0\right\rangle _{c,4}(\left| 0\right\rangle _A+\left|
1\right\rangle _A).
\end{equation}
Here and below a normalized factor is omitted for simplicity.

Step (ii) Adjust the level spacings of qubits ($1,2,3$) such that each of
these qubits is decoupled from its own cavity, and adjust the level spacings
of qubit $A$ to bring the $\left| 1\right\rangle \leftrightarrow \left|
2\right\rangle $ transition of this qubit dispersively coupled to the mode
of each of cavities $1,2,$and $3$ (i.e., $\Delta _{c,i}=\omega _{21}-\omega
_{c,i}\gg g_i$ for cavity $i$ with $i=1,2,3$) while the transition between
any other two levels of qubit $A$ is far-off resonant with (decoupled from)
the mode of each of cavities $1,2,3$ [Fig.~5(b)]. After an interaction time $%
t$, the state (17) changes to
\begin{eqnarray}
&&\ \ \left\{ \prod_{i=1}^3\left| 0\right\rangle _i\otimes \left[
\prod_{i=1}^3(\left| 0\right\rangle _{c,i}-i\left| 1\right\rangle
_{c,i})\left| 0\right\rangle _A\right. \right.  \nonumber \\
&&\ \ \left. \left. +\prod_{i=1}^3\left( \left| 0\right\rangle
_{c,i}-ie^{ig_i^2t/\Delta _{c,i}}\left| 1\right\rangle _{c,i}\right) \left|
1\right\rangle _A\right] \right\} \otimes \left| 0\right\rangle _4\left|
0\right\rangle _{c,4}.
\end{eqnarray}
With a choice of $\frac{g_1^2}{\Delta _{c,1}}=\frac{g_2^2}{\Delta _{c,2}}=%
\frac{g_3^2}{\Delta _{c,3}}$ and for $g_i^2\tau /\Delta _{c,i}=\pi ,$ we
obtain from Eq.~(18)
\begin{eqnarray}
&&\ \ \left\{ \prod_{i=1}^3\left| 0\right\rangle _i\otimes \left[
\prod_{i=1}^3(\left| 0\right\rangle _{c,i}-i\left| 1\right\rangle
_{c,i})\left| 0\right\rangle _A\right. \right.  \nonumber \\
&&\ \ \left. \left. +\prod_{i=1}^3\left( \left| 0\right\rangle
_{c,i}+i\left| 1\right\rangle _{c,i}\right) \left| 1\right\rangle _A\right]
\right\} \otimes \left| 0\right\rangle _4\left| 0\right\rangle _{c,4}.
\end{eqnarray}

Step (iii) Adjust the level spacings of qubit $A$ to its original
configuration [Fig.~5(a)] such that this qubit is decoupled from each
cavity. Then, adjust the level spacings of qubits ($1,2,3$) to bring the $%
\left| 0\right\rangle \leftrightarrow \left| 1\right\rangle $ transition of
qubit $i$ resonant with the mode of cavity $i$ ($i=1,2,3$) for an
interaction time $t_i=\pi /(2g_{r,i}),$ such that the state $\left|
0\right\rangle _i\left| 1\right\rangle _{c,i}$ is transformed to $-i\left|
1\right\rangle _i\left| 0\right\rangle _{c,i}$ while the state $\left|
0\right\rangle _i\left| 0\right\rangle _{c,i}$ remains unchanged. After this
step of operation, the state (19) becomes
\begin{equation}
\left[ \prod_{i=1}^3\left( \left| 0\right\rangle _i-\left| 1\right\rangle
_i\right) \left| 0\right\rangle _A+\prod_{i=1}^3\left( \left| 0\right\rangle
_i+\left| 1\right\rangle _i\right) \left| 1\right\rangle _A\right] \otimes
\left| 0\right\rangle _4\prod_{i=1}^4\left| 0\right\rangle _{c,i}.
\end{equation}
The result (20) shows that after the operation of this step, the qubit
system is disentangled from the cavities but the qubits ($1,2,3$) are
entangled with qubit $A.$

Step (iv) Adjust the level spacings of the qubits ($1,2,3$) such that these
qubits are decoupled from their cavities. Then, adjust the level spacings of
qubit $A$ such that the $\left| 1\right\rangle \leftrightarrow \left|
2\right\rangle $ transition of qubit $A$ is resonant with the mode of cavity
$4$ [Fig.~5(c)]. After an interaction time $t_A=\pi /(2g_{r,A}),$ the state $%
\left| 1\right\rangle _A\left| 0\right\rangle _{c,4}$ is transformed to $%
-i\left| 0\right\rangle _A\left| 1\right\rangle _{c,4}$ while the state $%
\left| 0\right\rangle _A\left| 0\right\rangle _{c,4}$ remains unchanged.
Here and below, $g_{r,A}$ is the resonant coupling constant of qubit $A$
with cavitiy $4$ while $g_{r,4}$ is the resonant coupling constant of qubit $%
4$ with cavity $4.$ After the operation of this step, the state (20) changes
to
\begin{equation}
\left[ \prod_{i=1}^3\left( \left| 0\right\rangle _i-\left| 1\right\rangle
_i\right) \left| 0\right\rangle _{c,4}-i\prod_{i=1}^3\left( \left|
0\right\rangle _i+\left| 1\right\rangle _i\right) \left| 1\right\rangle
_{c,4}\right] \otimes \left| 0\right\rangle _A\left| 0\right\rangle
_4\prod_{i=1}^3\left| 0\right\rangle _{c,i}.
\end{equation}

Step (v) Adjust the level spacings of the qubit $4$ such that this qubit is
now resonant with the mode of cavity $4$ for an interaction time $t_4=\pi
/(2g_{r,4})$ to transform the state $\left| 0\right\rangle _4\left|
1\right\rangle _{c,4}$ to $-i\left| 1\right\rangle _4\left| 0\right\rangle
_{c,4}$ while the state $\left| 0\right\rangle _4\left| 0\right\rangle
_{c,4} $ remains unchanged. As a result, the state (21) becomes

\begin{equation}
\left[ \prod_{i=1}^3\left( \left| 0\right\rangle _i-\left| 1\right\rangle
_i\right) \left| 0\right\rangle _4-\prod_{i=1}^3\left( \left| 0\right\rangle
_i+\left| 1\right\rangle _i\right) \left| 1\right\rangle _4\right] \otimes
\left| 0\right\rangle _A\prod_{i=1}^3\left| 0\right\rangle _{c,i},
\end{equation}
where $\left| -\right\rangle _i=\left| 0\right\rangle _i-\left|
1\right\rangle _i$ and $\left| +\right\rangle _i=\left| 0\right\rangle
_i+\left| 1\right\rangle _i.$ Note that after the operation of this step,
the level spacings of qubit $4$ need to be adjusted to have qubit $4$ to be
decoupled from cavity $4$.

From Eq.~(22), one can see that after the above operations, the qubits ($%
1,2,3,4$) are prepared in an GHZ state while each cavity returns to its
original vacuum state.

We should mention that because the level spacing between the two levels $%
\left| 1\right\rangle $ and $\left| 2\right\rangle $ of qubit $A$ in Fig.
5(a) is set to be greater than that in Fig. 5(b), qubit $A$ remains
off-resonant with any of the three resonators $1$, $2,$ and $3$ during
tuning the level structure of qubit $A$ from Fig. 5(a) to Fig. 5(b). Also,
when tuning the level spacings of qubit $A$ from Fig.~5(a) to Fig.~5(b), the
detuning between the $\left| 1\right\rangle \leftrightarrow $ $\left|
2\right\rangle $ transition frequency of qubit $A$ and the frequency of
resonator $4$ increases, and thus qubit $A$ is decoupled from resonator $4$
during the operations of steps (i)$\sim $(iv) above.

During the above GHZ-state preparation for the four qubits ($1,2,3,4$), the
other qubits in each cavity, which are represented by the grey dots in
Fig.~2(b), are decoupled from the cavity mode by prior adjustment of their
level spacings.

\begin{center}
\textbf{B. Fidelity}
\end{center}

Let us now study the fidelity of the entanglement preparation above. We note
that since the qubit-cavity resonant interaction or/and the qubit-pulse
resonant interaction are used in steps (i), (iii), (iv) and (v), these steps
can be completed within a very short time (e.g., by increasing the resonant
atom-cavity coupling constants), such that the dissipation of the qubits and
the cavities is negligibly small. In this case, the dissipation of the
system would appear in the operation of step (ii) due to the use of the
qubit-cavity dispersive interaction.

\begin{figure}[tbp]
\begin{center}
\includegraphics[bb=0 0 509 332, width=9.0 cm, clip]{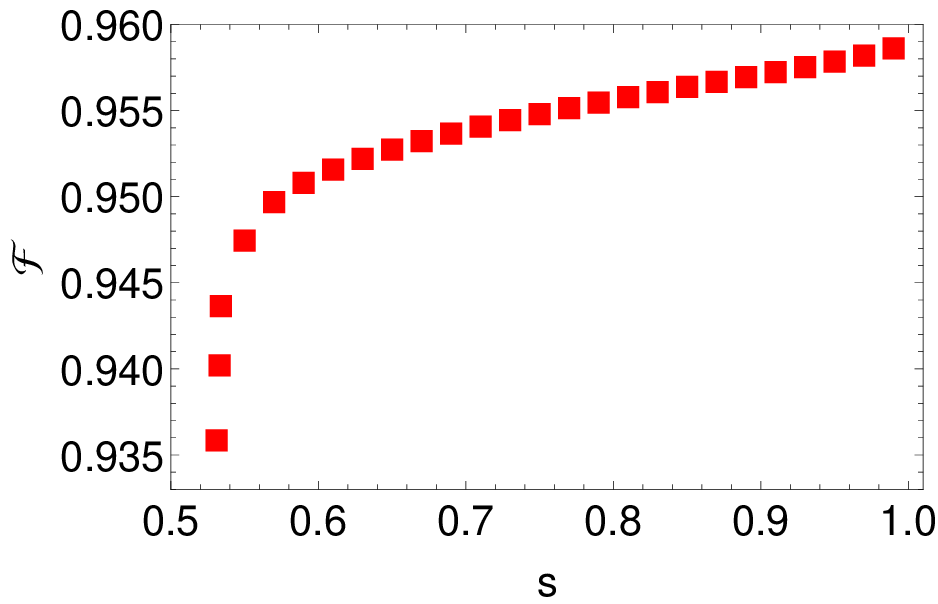} %
\vspace*{-0.08in}
\end{center}
\caption{(Color online) Fidelity versus $s$. The parameters used in the
numerical calculation are $\gamma _\varphi ^{-1}=(\gamma _\varphi ^{\prime
})^{-1}=(\gamma _\varphi ^{\prime \prime })^{-1}=5$ $\mu $s, $\gamma
^{-1}=25 $ $\mu $s, $(\gamma ^{\prime })^{-1}=200$ $\mu $s, $(\gamma
^{\prime \prime })^{-1}=50$ $\mu $s, $\kappa _1^{-1}=\kappa _2^{-1}=\kappa
_3^{-1}=20$ $\mu $s, $\Delta _{c,1}=10g_1,$ and $g_1/2\pi =100$ MHz.}
\label{fig:6}
\end{figure}

By defining $\Delta _{c,3}-\Delta _{c,2}=\Delta _{c,2}-\Delta _{c,1}=s\Delta
_{c,1},$ we have $\omega _{c,2}=\omega _{c,1}-s\Delta _{c,1}$ and $\omega
_{c,3}=\omega _{c,1}-2s\Delta _{c,1}.$ In addition, according to $%
g_1^2/\Delta _{c,1}=g_2^2/\Delta _{c,2}=g_3^2/\Delta _{c,3},$ we have $g_2=%
\sqrt{1+s}g_1$ and $g_3=\sqrt{1+2s}g_1.$ For the choice of $\gamma _\varphi
^{-1}=(\gamma _\varphi ^{\prime })^{-1}=(\gamma _\varphi ^{\prime \prime
})^{-1}=5$ $\mu $s, $\gamma ^{-1}=25$ $\mu $s, $(\gamma ^{\prime })^{-1}=200$
$\mu $s, $(\gamma ^{\prime \prime })^{-1}=50$ $\mu $s, $\kappa
_1^{-1}=\kappa _2^{-1}=\kappa _3^{-1}=20$ $\mu $s, $\Delta _{c,1}=10g_1,$
and $g_1/2\pi =100$ MHz, the fidelity versus the parameter $s$ is shown in
Fig.~6, from which one can see that a high fidelity $\sim 96\%$ can be
achieved when $s=1,$ which corresponds to the case of $g_2/2\pi \sim 141$
MHz and $g_3\sim 173$ MHz. In the following we consider the case of $s=1.$
Without loss of generality, assume that the $\left| 1\right\rangle
\leftrightarrow \left| 2\right\rangle $ transition frequency of qubit $A$ is
$\nu _0\sim 8.5$ GHz, and thus the frequency of cavity $1,$ the frequency of
cavity $2$ and the frequency of cavity $3$ are $\nu _{c,1}\sim 7.5$ GHz, $%
\nu _{c,2}\sim 6.5$ GHz and $\nu _{c,3}\sim 5.5$ GHz, respectively. For the
cavity frequencies chosen here and for the $\kappa _1^{-1},\kappa
_2^{-1},\kappa _3^{-1}$ used in our numerical calculation, the required
quality factors for cavities $1,2,$and $3$ are $Q_1\sim 9.4\times 10^5,$ $%
Q_2\sim 8.2\times 10^5,$ and $Q_3\sim 6.9\times 10^5,$ respectively.
Finally, it is noted that since only resonant interaction of qubit $A$ with
cavity $4$ is involved during the above operations, the requirement for
cavity $4$ is greatly reduced when compared with cavities $1,$ $2,$ and $3.$
Our analysis given here shows that preparation of a GHZ entangled state of
four qubits located at four cavities is possible within the present circuit
cavity QED technique.

\begin{center}
\textbf{C. Preparation of GHZ states of multiple qubits located at }$n$\textbf{\ cavities}
\end{center}

One can easily verify that \textrm{in principle} by using a superconducting
qubit coupled to $n$ cavities, $n$ qubits ($1,2,...,n$) initially in the
state $\prod_{i=1}^{n-1}\left| +\right\rangle _i\otimes \left|
0\right\rangle _n$, which are respectively located in the different $n$
cavities, can be prepared in an entangled GHZ state $\prod_{i=1}^{n-1}\left|
-\right\rangle _i\left| 0\right\rangle _n-\prod_{i=1}^n\left| +\right\rangle
_i\left| 1\right\rangle _n$ by using the same procedure described above.

Furthermore, based on the prepared GHZ state of $n$ qubits ($1,2,...,n$),
all other qubits (not entangled initially) in the cavities can be entangled
with the GHZ-state qubits ($1,2,...,n$), through \textrm{intra-cavity}
controlled-NOT (CNOT) operations on the qubits in each \textrm{cavity by}
using the GHZ-state qubit in each cavity (i.e., qubit $1,$ $2,...,$ or $n$)
as the control while the other qubits as the targets. To see this clearly,
let us consider Fig.~2(b), where the three qubits in cavity $i$ ($i=1,2,3,4$%
)\ are the black-dot qubit $i$ and the two grey-dot qubits, labelled as
qubits $i2$ and $i3$ here. Suppose that the four black-dot qubits ($1,2,3,4$%
) (i.e., the GHZ-state qubits) were prepared in the GHZ state of Eq.~(17),
and each grey-dot qubit is initially in the state $\left| +\right\rangle $.
By \textrm{performing CNOT on} \textrm{various qubit pairs }in each cavity,
\textit{i.e}., $C_{i,i2}$ \textrm{and }$C_{i,i3}$\textrm{\ on the qubit
pairs (}$i,i2)$\textrm{\ and (}$i,i3$\textrm{)} for cavity $i$, one can have
all qubits in the four cavities (both black-dot and grey-dot qubits)
prepared in a GHZ state $\prod_{i=1}^4\left| -\right\rangle _i\left|
-\right\rangle _{i2}\left| -\right\rangle _{i3}-\prod_{i=1}^4\left|
+\right\rangle _i\left| +\right\rangle _{i2}\left| +\right\rangle _{i3}$.
Here, $C_{i,i2},$ defined in the basis $\{\left| +\right\rangle _i\left|
+\right\rangle _{i2},\left| -\right\rangle _i\left| +\right\rangle
_{i2},\left| +\right\rangle _i\left| -\right\rangle _{i2},$ $\left|
-\right\rangle _i\left| -\right\rangle _{i2}\},$ represents a CNOT with
qubit $i$ (the GHZ-state qubit) as the control while qubit $i2$ as the
target, which results in the transformation $\left| -\right\rangle _i\left|
+\right\rangle _{i2}\rightarrow $ $\left| -\right\rangle _i\left|
-\right\rangle _{i2}$ while leaves the state $\left| +\right\rangle _i\left|
+\right\rangle _{i2}$ unchanged. A similar definition \textrm{applies} to $%
C_{i,i3.}$ Alternatively, using the prepared GHZ state of $n$ qubits ($%
1,2,...,n$), one can have all other qubits in the cavities to be entangled
with the GHZ-state qubits ($1,2,...,n$), by performing an \textrm{%
intra-cavity} multiqubit CNOT with the GHZ-state qubit (control qubit)
simultaneously controlling all other qubits (target qubits) in each cavity
[23].

Experimentally, it has been demonstrated successfully on circuits consisting
up to 128 flux qubits that crosstalk from control circuitry can be
essentially eliminated and/or corrected by practicing proper circuit designs
and developing corresponding multilayer fabrication processes [58]. Hence,
frequency crowding for multiple qubits in one resonator, and control of
large numbers of qubits do not present a fundamental and/or practical
problem for the proposed protocol.

\begin{figure}[tbp]
\begin{center}
\includegraphics[bb=56 311 551 630, width=8.6 cm, clip]{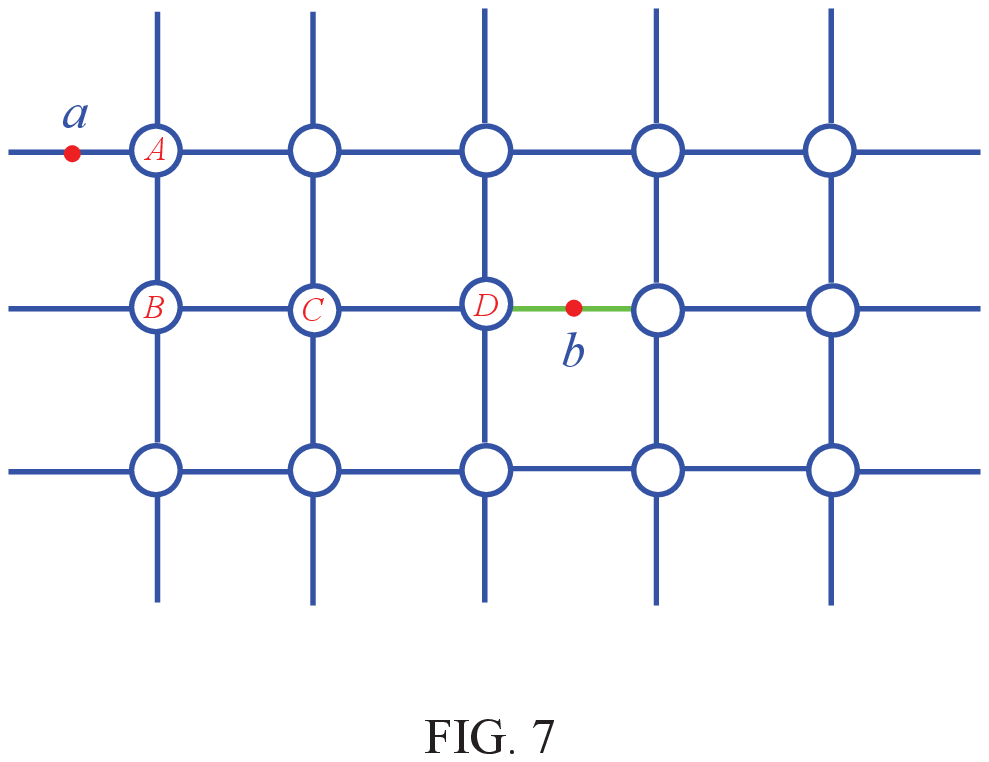} %
\vspace*{-0.08in}
\end{center}
\caption{(Color online) Two-dimensional linear network of resonators and
qubits. A short line represents a resonator and each \textrm{circle}
represents a coupler qubit. The two red dots represent qubits $a$ and $b$.
The coupler qubits $A,B$ and $C$ are used to transfer information stored in
qubit $a$ to the coupler qubit $D$. They are also used to transfer
information of the coupler qubit $D$ back to qubit $a$ after a quantum
operation is performed on the coupler qubit $D$ and qubit $b$, which
interact with each other through a resonator (i.e., the green short line).}
\label{fig:7}
\end{figure}

\begin{center}
\textbf{V. POSSIBILITY OF A TWO-DIMENSIONAL QUANTUM NETWORK}
\end{center}

The four resonators coupled by a coupler superconducting qubit may be used
as a basic circuit block to build a two-dimensional (2D) quantum network for
quantum information processing, as depicted in Fig.~7. In this network, for
any two qubits coupled or connected by a resonator (e.g., qubits $a$ and $A$%
, qubits $A$ and $B$, and so on), quantum operations can be performed on
them directly because the two qubits can interact with each other, mediated
by their shared resonator. In addition, for any two qubits located at
different cavities or resonators, quantum operations can be performed
through information transfer. To see this, let us consider two distant
qubits $a$ and $b$ in the network [Fig.~7)]. To perform a quantum operation
on the two qubits $a$ and $b,$ one can do as follows. First, transfer the
quantum information stored in qubit $a$ to the coupler qubit $D$ via a
transfer sequence $a\rightarrow A\rightarrow B\rightarrow C\rightarrow D$,
(ii) perform the quantum operation on the coupler qubit $D$ and qubit $b,$
and then (iii) transfer the quantum information of the coupler qubit $D$
back to qubit $a$ through a transfer sequence $D\rightarrow C\rightarrow
B\rightarrow A\rightarrow a$. In this way, the quantum operation is
performed on the two distant qubits $a$ and $b$ indirectly. It should be
mentioned that to perform a quantum operation on two qubits at different
cavities, the intermediate coupler qubits (e.g., qubits $A$, $B$, and $C$
for the example given here) need to be initially prepared in the ground
state $\left| 0\right\rangle $ as required by quantum information transfer
(e.g., this can be see from the state transformation $\left( \alpha \left|
0\right\rangle _a+\beta \left| 1\right\rangle _a\right) \left|
0\right\rangle _A\rightarrow \left| 0\right\rangle _a\left( \alpha \left|
0\right\rangle _A+\beta \left| 1\right\rangle _A\right) $ for the
information transfer from qubit $a$ to the coupler qubit $A$).

An architecture for quantum computing based on superconducting circuits,
where on-chip planar microwave resonators are arranged in a two-dimensional
grid with a qubit sitting at each intersection, was previously presented
[59]. However, our present proposal is different from theirs in the
following. For the architecture in Ref.~[59], each qubit at an intersection
is coupled to two cavity modes, i.e., one cavity mode belongs to a
horizontal cavity built on the top layer while the other cavity mode belongs
to a vertical cavity built at a second layer at the bottom. In contrast, in
our case, as shown in Fig.~7, all resonators and coupler qubits are arranged
in the same plane, which is relatively easy to be implemented in experiments.

Finally, Ref. [60] analyzes the performance of the Resonator/zero-Qubit
(RezQu) architecture in which the qubits are complemented with memory
resonators and coupled via a resonator bus. We note that in Ref. [60], the
memory resonators are coupled via a common resonator bus, while in our
proposal the cavities are coupled via a coupler qubit. Hence, our
architecture is quite different from the one in [60].

We remark that many details on possible scalability of the protocol
(including quantum error correction) need to be addressed. However, this
requires a lengthy and complex analysis, which is beyond the scope of the
present work. We would like to leave them as open questions to be addressed
in future work.

\begin{center}
\textbf{VI. CONCLUSION}
\end{center}

We have proposed a method for creating four-resonator entangled coherent
states and preparing a GHZ state of four qubits in four cavities, by using a
superconducting qubit as the coupler. In principle, this proposal can be
extended to create entangled coherent states of $n$ resonators and to
prepare GHZ states of $n$ qubits distributed over $n$ cavities in a network,
with the same operational steps and the operation time as those of the
four-resonator case described above. This proposal is quite general, which
can be applied to other types of physical qubit systems with three levels,
such as quantum dots and NV centers coupled to \textrm{cavities}. Finally,
it is noted that the four resonators coupled by a coupler qubit can be used
as a basic circuit block to build a two-dimensional quantum network, which
may be useful for scalable quantum information processing.

\begin{center}
\textbf{ACKNOWLEDGMENTS}
\end{center}

S. Han was supported in part by DMEA. C.P. Yang was supported in part by the
National Natural Science Foundation of China under Grant No. 11074062, the
Zhejiang Natural Science Foundation under Grant No. Y6100098, the Open Fund
from the SKLPS of ECNU, and the funds from Hangzhou Normal University. Q.P.
Su was supported by the National Natural Science Foundation of China under
Grant No. 11147186. S. B. Zheng was supported in part by the National
Fundamental Research Program Under Grant No. 2012CB921601, National Natural
Science Foundation of China under Grant No. 10974028, the Doctoral
Foundation of the Ministry of Education of China under Grant No.
20093514110009, and the Natural Science Foundation of Fujian Province under
Grant No. 2009J06002.

\end{document}